\documentclass[dvips]{article}
\usepackage{amssymb}
\usepackage{amsmath}
\usepackage{rotating}
\DeclareSymbolFont{ppa}{OT1}{ppl}{m}{it}
\DeclareMathSymbol{\vv}{\mathalpha}{ppa}{'166}

\begin{document}

\newcommand{\dd}{\,{\rm d}}
\newcommand{\ie}{{\it i.e.},\,}
\newcommand{\etal}{{\it et al.\ }}
\newcommand{\eg}{{\it e.g.},\,}
\newcommand{\cf}{{\it cf.\ }}
\newcommand{\vs}{{\it vs.\ }}
\newcommand{\zdot}{\makebox[0pt][l]{.}}
\newcommand{\up}[1]{\ifmmode^{\rm #1}\else$^{\rm #1}$\fi}
\newcommand{\dn}[1]{\ifmmode_{\rm #1}\else$_{\rm #1}$\fi}
\newcommand{\upd}{\up{d}}
\newcommand{\uph}{\up{h}}
\newcommand{\upm}{\up{m}}
\newcommand{\ups}{\up{s}}
\newcommand{\arcd}{\ifmmode^{\circ}\else$^{\circ}$\fi}
\newcommand{\arcm}{\ifmmode{'}\else$'$\fi}
\newcommand{\arcs}{\ifmmode{''}\else$''$\fi}
\newcommand{\MS}{{\rm M}\ifmmode_{\odot}\else$_{\odot}$\fi}
\newcommand{\RS}{{\rm R}\ifmmode_{\odot}\else$_{\odot}$\fi}
\newcommand{\LS}{{\rm L}\ifmmode_{\odot}\else$_{\odot}$\fi}
\newcommand{\feh}{\hbox{$ [{\rm Fe}/{\rm H}]$}}

\newcommand{\Abstract}[2]{{\footnotesize\begin{center}ABSTRACT\end{center}
\vspace{1mm}\par#1\par
\noindent
{~}{\it #2}}}

\newcommand{\TabCap}[2]{\begin{center}\parbox[t]{#1}{\begin{center}
  \small {\spaceskip 2pt plus 1pt minus 1pt T a b l e}
  \refstepcounter{table}\thetable \\[2mm]
  \footnotesize #2 \end{center}}\end{center}}

\newcommand{\TableSep}[2]{\begin{table}[p]\vspace{#1}
\TabCap{#2}\end{table}}

\newcommand{\FigCap}[1]{\footnotesize\par\noindent Fig.\  %
  \refstepcounter{figure}\thefigure. #1\par}

\newcommand{\TableFont}{\footnotesize}
\newcommand{\TableFontIt}{\ttit}
\newcommand{\SetTableFont}[1]{\renewcommand{\TableFont}{#1}}

\newcommand{\MakeTable}[4]{\begin{table}[htb]\TabCap{#2}{#3}
  \begin{center} \TableFont \begin{tabular}{#1} #4
  \end{tabular}\end{center}\end{table}}

\newcommand{\MakeTableSep}[4]{\begin{table}[p]\TabCap{#2}{#3}
  \begin{center} \TableFont \begin{tabular}{#1} #4
  \end{tabular}\end{center}\end{table}}

\newenvironment{references}%
{
\footnotesize \frenchspacing
\renewcommand{\thesection}{}
\renewcommand{\in}{{\rm in }}
\renewcommand{\AA}{Astron.\ Astrophys.}
\newcommand{\AAS}{Astron.~Astrophys.~Suppl.~Ser.}
\newcommand{\ApJ}{Astrophys.\ J.}
\newcommand{\ApJS}{Astrophys.\ J.~Suppl.~Ser.}
\newcommand{\ApJL}{Astrophys.\ J.~Letters}
\newcommand{\AJ}{Astron.\ J.}
\newcommand{\IBVS}{IBVS}
\newcommand{\PASJ}{PASJ}
\newcommand{\PASP}{P.A.S.P.}
\newcommand{\Acta}{Acta Astron.}
\newcommand{\MNRAS}{MNRAS}
\renewcommand{\and}{{\rm and }}
\section{{\rm REFERENCES}}
\sloppy \hyphenpenalty10000
\begin{list}{}{\leftmargin1cm\listparindent-1cm
\itemindent\listparindent\parsep0pt\itemsep0pt}}%
{\end{list}\vspace{2mm}}

\def\TYLDA{~}
\newlength{\DW}
\settowidth{\DW}{0}
\newcommand{\dw}{\hspace{\DW}}

\newcommand{\refitem}[5]{\item[]{#1} #2%
\def\REFARG{#3}\ifx\REFARG\TYLDA\else, {\it#3}\fi
\def\REFARG{#4}\ifx\REFARG\TYLDA\else, {\bf#4}\fi
\def\REFARG{#5}\ifx\REFARG\TYLDA\else, {#5}\fi.}

\newcommand{\Section}[1]{\section{#1}}
\newcommand{\Subsection}[1]{\subsection{#1}}
\newcommand{\Acknow}[1]{\par\vspace{5mm}{\bf Acknowledgments.} #1}
\pagestyle{myheadings}

\newfont{\bb}{ptmbi8t at 12pt}
\newcommand{\xrule}{\rule{0pt}{2.5ex}}
\newcommand{\xxrule}{\rule[-1.8ex]{0pt}{4.5ex}}
\def\thefootnote{\fnsymbol{footnote}}

\begin{center}
{\Large\bf Identification of V735 Sgr as an Active\\ Herbig Ae/Be
Object\footnote{Based on observations obtained with the 1.3-m Warsaw telescope
and the 6.5-m Magellan Baade telescope at the Las Campanas Observatory
of the Carnegie Institution for Science under the CNTAC program CN2018A-102.}}
\vskip1cm
{\bf
P.~~P~i~e~t~r~u~k~o~w~i~c~z$^1$,~~F.~~D~i~~M~i~l~l~e$^2$,~~R.~~A~n~g~e~l~o~n~i$^{3,4}$\\
A.~U~d~a~l~s~k~i$^1$\\}
\vskip3mm
{
$^1$ Warsaw University Observatory, Al. Ujazdowskie 4, 00-478 Warszawa, Poland\\
e-mail: pietruk@astrouw.edu.pl\\
$^2$ Las Campanas Observatory, Casilla 601, La Serena, Chile\\
$^3$ Departamento de F\'isica y Astronom\'ia, Universidad de La Serena, Av. Cisternas 1200 Norte, La Serena, Chile\\
$^4$ Instituto de Investigaci\'on Multidisciplinar en Ciencia y Tecnolog\'ia, Universidad de La Serena, Av. Ra\'ul Bitr\'an 1305, La Serena, Chile\\
}
\end{center}

\Abstract{V735 Sgr was known as an enigmatic star with rapid brightness
variations. Long-term OGLE photometry, brightness measurements in infrared bands,
and recently obtained moderate resolution spectrum from the 6.5-m Magellan
telescope show that this star is an active young stellar object of Herbig Ae/Be
type.}

{Stars: variables: T Tauri, Herbig Ae/Be - Stars: individual: V735 Sgr}


\Section{Motivation}

V735 Sgr was found to be a variable object by Luyten (1937) who
analyzed photographic plates obtained in the course of the Bruce
Proper Motion Survey. Due to relatively large variations
in blue photographic brightness (14.2--15.5 mag) the object
was supposed to be an eruptive cataclysmic variable of Z Cam type
(2006 archival version of the Downes \etal 2001 catalog),
however no evident outburst or standstill had been observed.
A mistake made in the position of the variable source
(Vogt and Bateson 1982) complicated the identification of the true
nature of the star.


\Section{Photometric Observations}

V735 Sgr is located in the area monitored by the Optical Gravitational
Lensing Experiment (OGLE) since the beginning of the third phase
of the survey in 2001. The survey monitors the Galactic bulge,
Galactic disk, and Magellanic Clouds from Las Campanas Observatory,
Chile. Since March 2010 the project is in its fourth phase.
The OGLE-IV camera consists of 32 CCDs with a total field of view of about
1.4~deg$^2$. Currently, OGLE measures brightness in the Johnson $V$ and
Cousins $I$ passbands of over two billion stars of the Milky Way and the
Magellanic System, covering a total area of about 3600 deg$^2$.
Technical details on OGLE-IV, including data reduction, can be found
in Udalski \etal (2015).

OGLE observations confirm the source identification of V735 Sgr given by
Yoshida \etal (2002). The variable is located at the equatorial coordinates
$(\alpha,\delta)_{2000.0}$ $=(17\uph59\upm51\zdot\ups78,-29\arcd33\arcm55\zdot\arcs9)$
or at the Galactic coordinates $(l,b)$ $=(+1\zdot\arcd0372,$ $-2\zdot\arcd9953)$.
We identify it with OGLE detection BLG505.01.92650.
Fig.~1 shows finding charts in $VJHK_{\rm s}$ bands centered on the variable.
The $V$-band chart was cropped from the OGLE-IV reference frame for field
BLG505.01. $JHK_{\rm s}$-band charts come from images taken by the VISTA
Variables in the V\'ia Lactea (VVV) ESO public survey (Minniti \etal 2010).
In the upper panel of Fig.~2, we present full $I$-band light curve obtained
since the beginning of OGLE-III (2001) until the middle of the ninth bulge
season of OGLE-IV (2018). The rapid irregular behavior of V735 Sgr is well seen
in the middle and lower panels of Fig.~2. The full $I$-band light curve consists of
1304 data points collected during OGLE-III and 15529 data points collected during
OGLE-IV by the end of June 2018. Light curve in the $V$ band, not shown here,
consists of 7 measurements from OGLE-III and 182 measurements from OGLE-IV.
In the whole observed period 2001--2018, the $I$-band brightness varied between
12.18~mag and 15.54~mag, while $V$-band brightness ranged between 13.11 mag and
16.10 mag. This means that the full $I$-band and $V$-band amplitudes of V735 Sgr
were 3.36 mag and 2.99 mag, respectively. Measured $V-I$ color index fluctuated
between +0.58 and +1.06 mag. Mean $V$ and $I$-band magnitudes and mean $V-I$ color
during the OGLE-IV phase are the following: 14.74, 13.86, +0.88 mag, respectively.
The most rapid brightness change was recorded on the night from July 31 to August 1,
2015 when the star faded by 0.95 mag in 7.32~h (see lower panel of Fig.~2).
We report no periodic signal in the power spectrum of the variable.

\begin{figure}[htb!]
\centerline{\includegraphics[angle=0,width=124mm]{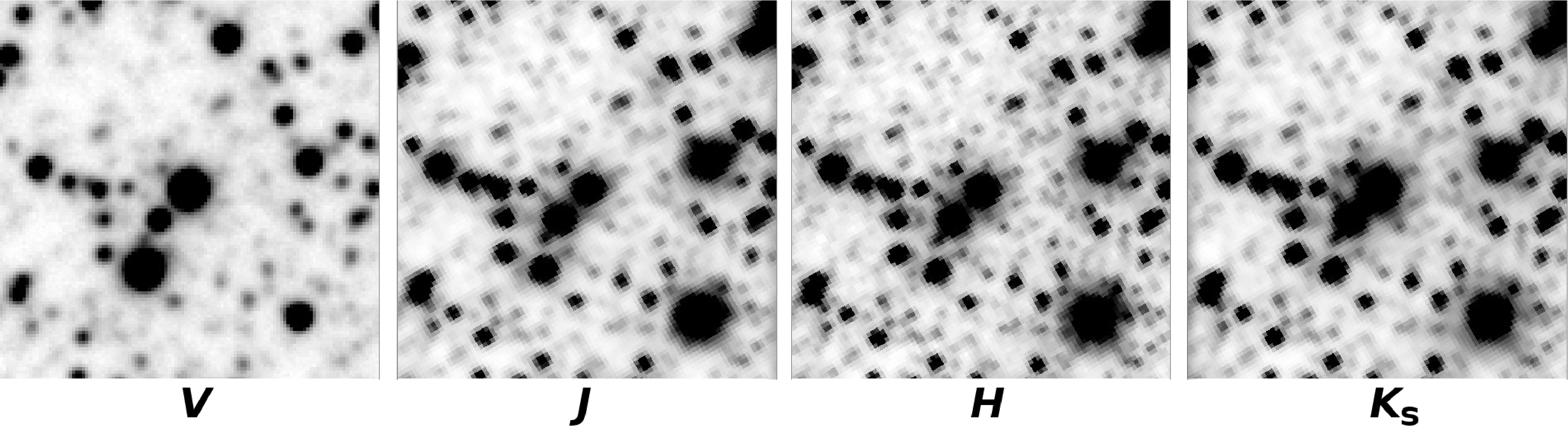}}
\FigCap{Finding charts in $V$, $J$, $H$, and $K_{\rm s}$ bands, each $30\arcs$
on a size, centered on V735 Sgr. North is up and East is to the left.
The comparison constant star used to demonstrate the infrared excess
in V735 Sgr (see Fig.~4) is located $7\zdot\arcs24$ roughly
South-East of the variable. The $V$-band chart was obtained with
the 1.3-m OGLE telescope, while the $JHK_{\rm s}$-band charts were
obtained in the course of the VVV survey on the 4.1-m VISTA telescope.}
\end{figure}

\begin{figure}[htb!]
\centerline{\includegraphics[angle=0,width=124mm]{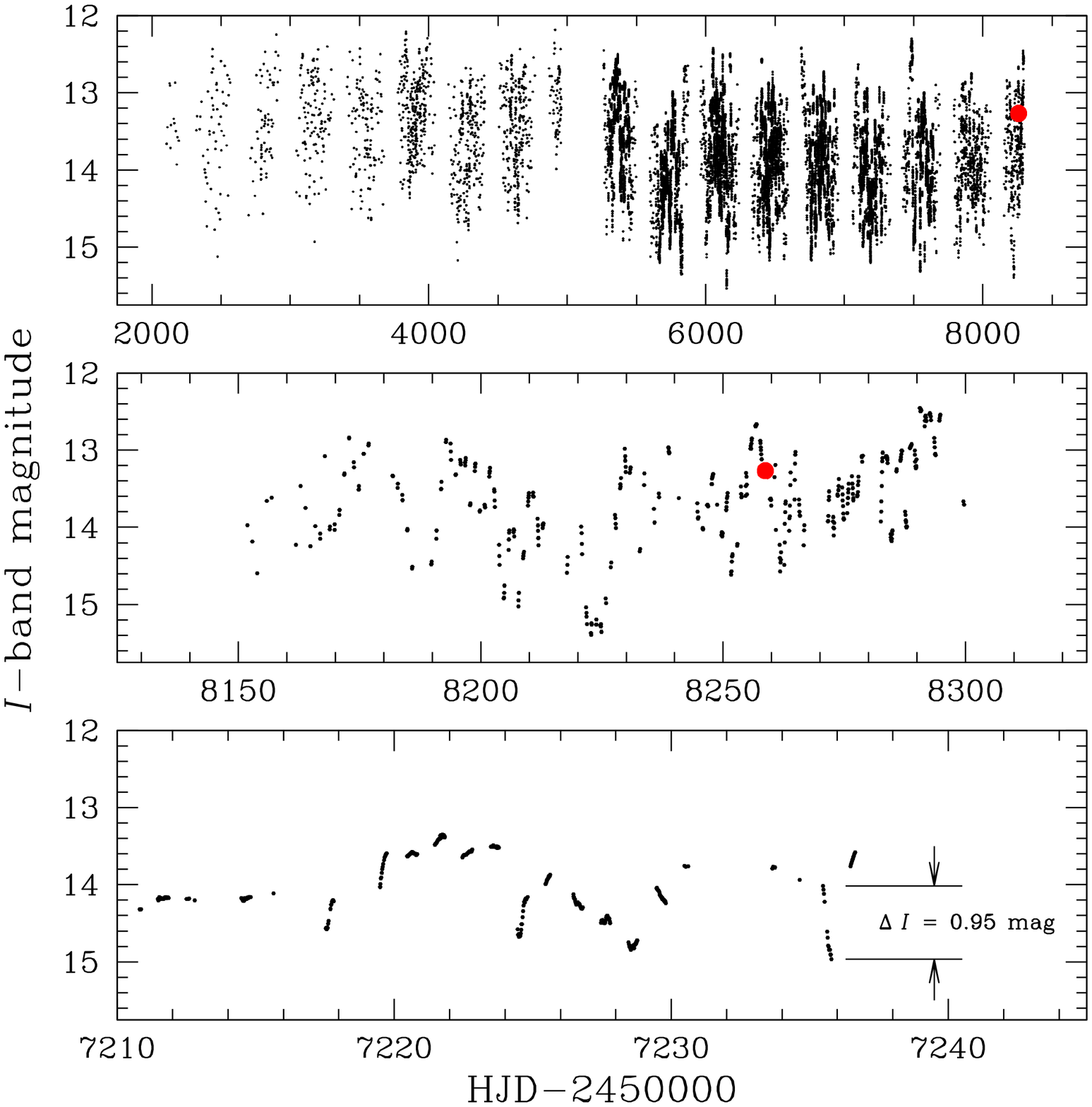}}
\FigCap{OGLE $I$-band light curve of variable V735 Sgr covering years
2001--2018 (upper panel), with a zoom on the data collected in the first
half of 2018 (middle panel) and on a part with the most rapid brightness
change that took place in 2015 (lower panel). The big red dot marks the moment
of the executed spectrum.}
\end{figure}

Based on the obtained time-series photometry we conclude that V735 Sgr exhibits permanent,
rapid, irregular variations. No outbursts or standstills observed over 18 years
clearly indicate that this object cannot be a dwarf nova of U Gem or Z Cam type.
Irregular fluctuations with the amplitude of about 3 mag are too large
for a novalike system. The recorded photometric behavior is characteristic
for a subgroup of young stellar objects of UX Ori type. This is additionally
confirmed with the observed color fluctuations and the presence of infrared excess.

The changing location of the variable in the $I$ \vs $V-I$ diagram is presented
in Fig.~3. The trend drawn by the star has a crescent shape, as it is observed
in some young stellar objects (\eg BF Ori, Evans \etal 1989, UX Ori, Grinin \etal 1994).
The so called blueing effect is well visible---with the decreasing brightness the
star gets redder but near minimum it becomes blue again mainly due to scattering
light by dust.

\begin{figure}[htb!]
\centerline{\includegraphics[angle=0,width=124mm]{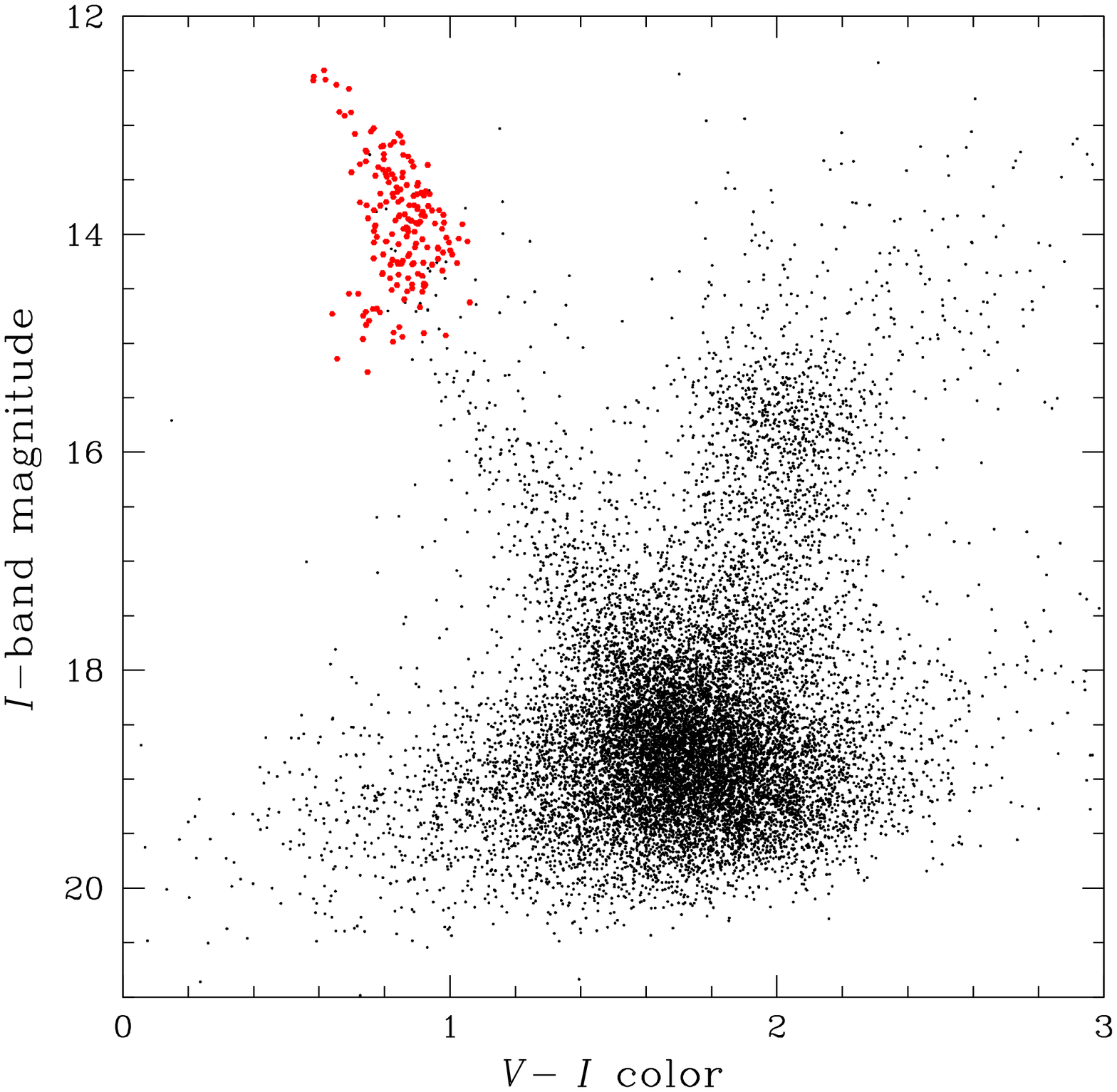}}
\FigCap{Color-magnitude diagram with marked positions of the variable star
V735 Sgr over years 2001--2018. The star draws a crescent shape
characteristic for young stellar objects. The background is formed of
stars located in the OGLE field BLG505.01 containing the variable.}
\end{figure}

\begin{figure}[htb!]
\centerline{\includegraphics[angle=0,width=124mm]{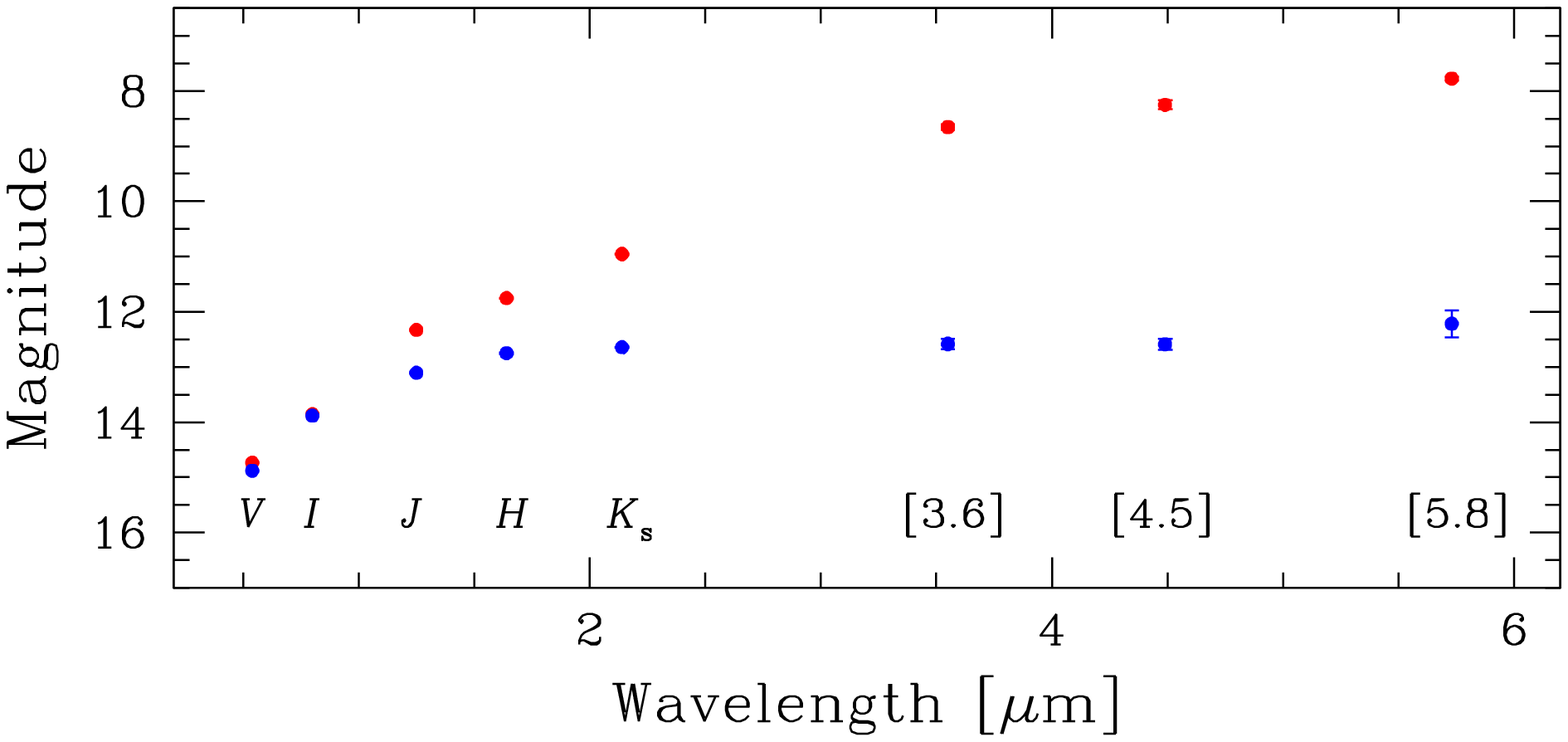}}
\FigCap{Brightness of V735 Sgr (in red) \vs neighbor constant star BLG505.01.92647
(in blue) in the optical and infrared regimes. The used $V$ and $I$-band
magnitudes are average values. Note a clear infrared excess in V735 Sgr
confirming its pre-main sequence nature.}
\end{figure}

In Fig.~4, we compare brightness from optical to mid-infrared regime of V735 Sgr
and a neighbor constant star BLG505.01.92647 located $7\zdot\arcs24$ South-East
of the variable. Near-infrared $JHK_{\rm s}$ magnitudes were taken from the
VVV survey, while mid-infrared data come from Spitzer/GLIMPSE 3D
catalog\footnote{https://irsa.ipac.caltech.edu/Missions/spitzer.html}.
The comparison star has very similar magnitudes
in the optical bands (that is also similar color) to average values for
the variable: $V=14.88$ and $I=13.88$ ($V-I=+1.00$ mag). V735 Sgr clearly
shows a strong excess in infrared, as it is observed in pre-main sequence stars.
However, a definitive classification of the variable star required
spectroscopic observations.


\Section{Spectroscopic Data and Results}

Two subsequent spectra of V735 Sgr were obtained with the 6.5-m Magellan
Baade telescope at Las Campanas Observatory in dark time on night
May 19/20, 2018. The night was not photometric due to the presence of some
thin cirrus clouds, but with stable seeing of about $0.6\arcs$. We used the
Magellan Echellette (MagE) spectrograph with a $1.0\arcs$ slit giving an
average spectral resolution of $R\approx4100$ over wavelength range
between 3200~\AA~and 10000~\AA. After two 300-s science exposures a ThAr lamp
was exposed for 15 s. Initial reductions, that is flat-field correction,
rectification of spectral orders, wavelength calibration and combination
of the two single exposures with rejection of cosmic rays, were performed
with the CarPy software (Kelson 2003) on the site. Using utilities provided
in the IRAF package\footnote{IRAF is distributed by the National Optical Astronomy
Observatory, which is operated by the Association of Universities for Research
in Astronomy, Inc., under a cooperative agreement with the National Science
Foundation.}, we normalized and also calibrated for flux each order separately.
For the calibration we used a spectrophotometric standard star, white dwarf
LTT7987 exposed in the dawn of the night ($2\times200$~s with the same instrument
setup). Then, all orders were combined to form one normalized spectrum and the other
one in units of flux. Due to low intensity at the blue end and the presence
of many atmospheric features in the near-infrared part of the spectrum,
we rely only on orders from 8 to 17, covering wavelength range 3500--8100~\AA.

Mean moment of the spectroscopic observation (HJD=2458258.71054)
is marked on the $I$-band light curve in Fig.~2. At that time the variable
had $I=13.27$~mag and was fading down from a local maximum of $I=12.66$~mag
for almost two days. Unfortunately, there is no OGLE observation in the $V$ band
from that night.

In Fig.~5, we present the flux-calibrated spectrum of V735 Sgr covering
the whole optical regime. The dominant stellar feature is the hydrogen Balmer
series with H$\alpha$ in strong emission. Such a spectrum is typical for a young
stellar object of spectral type A. Thus, V735 Sgr is an intermediate-mass 
object of Herbig Ae/Be type. The star seems to be mildly reddened,
likely due to its relative proximity ($855\pm44$~pc based on Gaia
Data Release 2, Gaia Collaboration \etal 2018)
and location in the direction of Baade's Window. The most prominent lines
are marked in the normalized spectrum in Fig.~6. Beside H$\alpha$ also
[O~{\scriptsize I}] and [S~{\scriptsize II}] lines are in emission.
The obtained spectrum reveals
the presence of He {\scriptsize I} and Mg {\scriptsize II} lines being
indicators of early type A of the underlying star. There are absorption
lines of singly-ionized metals such as Ca {\scriptsize I} and
Fe {\scriptsize I} and the multiplet Fe {\scriptsize II} (42) containing
relatively strong lines at 4924\AA, 5018\AA,~and 5169\AA. All of these
features are specific for young stellar objects.

\begin{figure}[htb!]
\centerline{\includegraphics[angle=0,width=124mm]{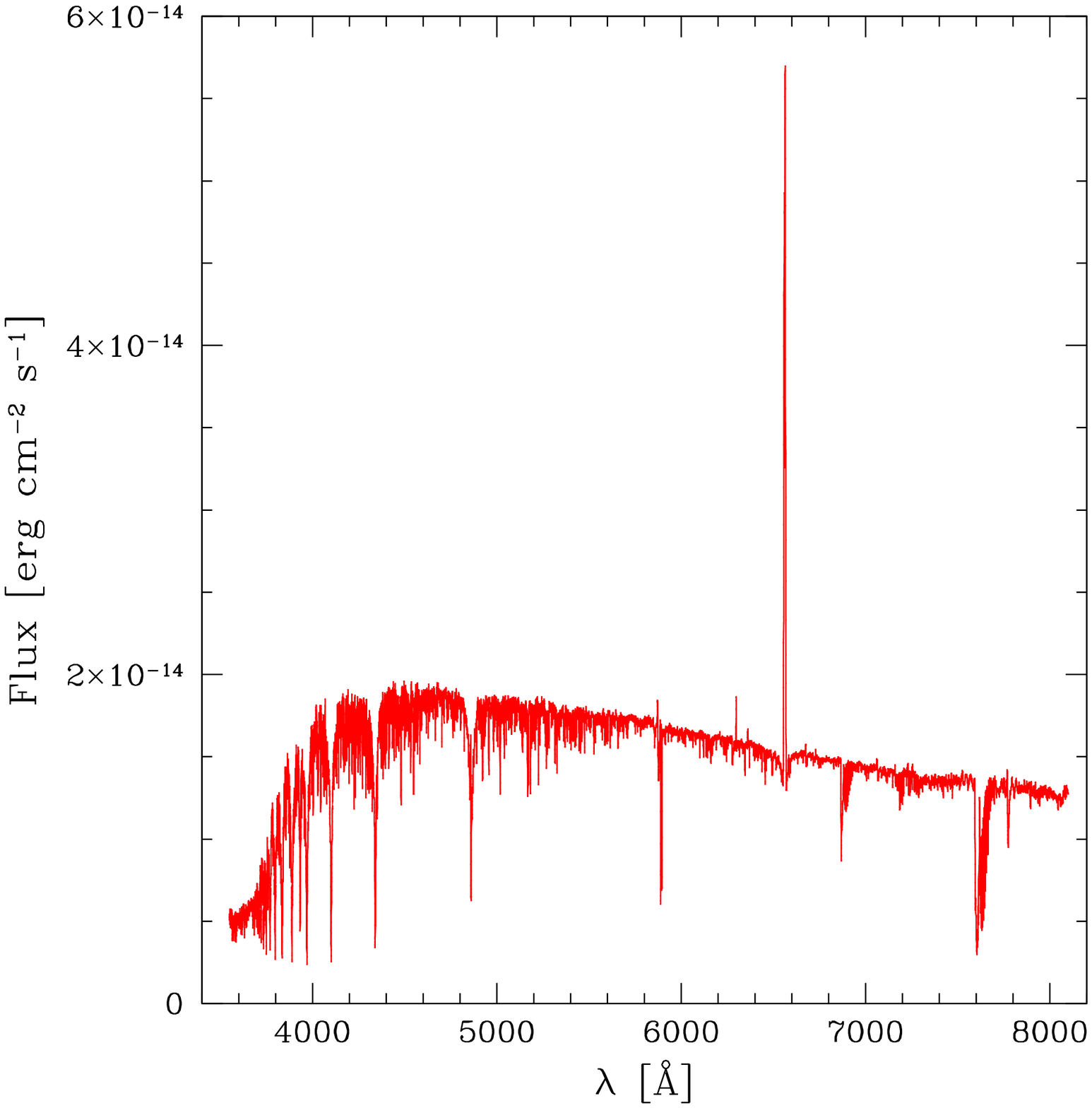}}
\FigCap{Flux-calibrated spectrum of V735 Sgr. Note the presence of the
dominant hydrogen Balmer series with a strong emission in H$\alpha$ line
characteristic for young stellar objects of spectral type A.
The spectrum is not corrected for atmospheric absorption features.}
\end{figure}

\begin{figure}[htb!]
\centerline{\includegraphics[angle=0,width=124mm]{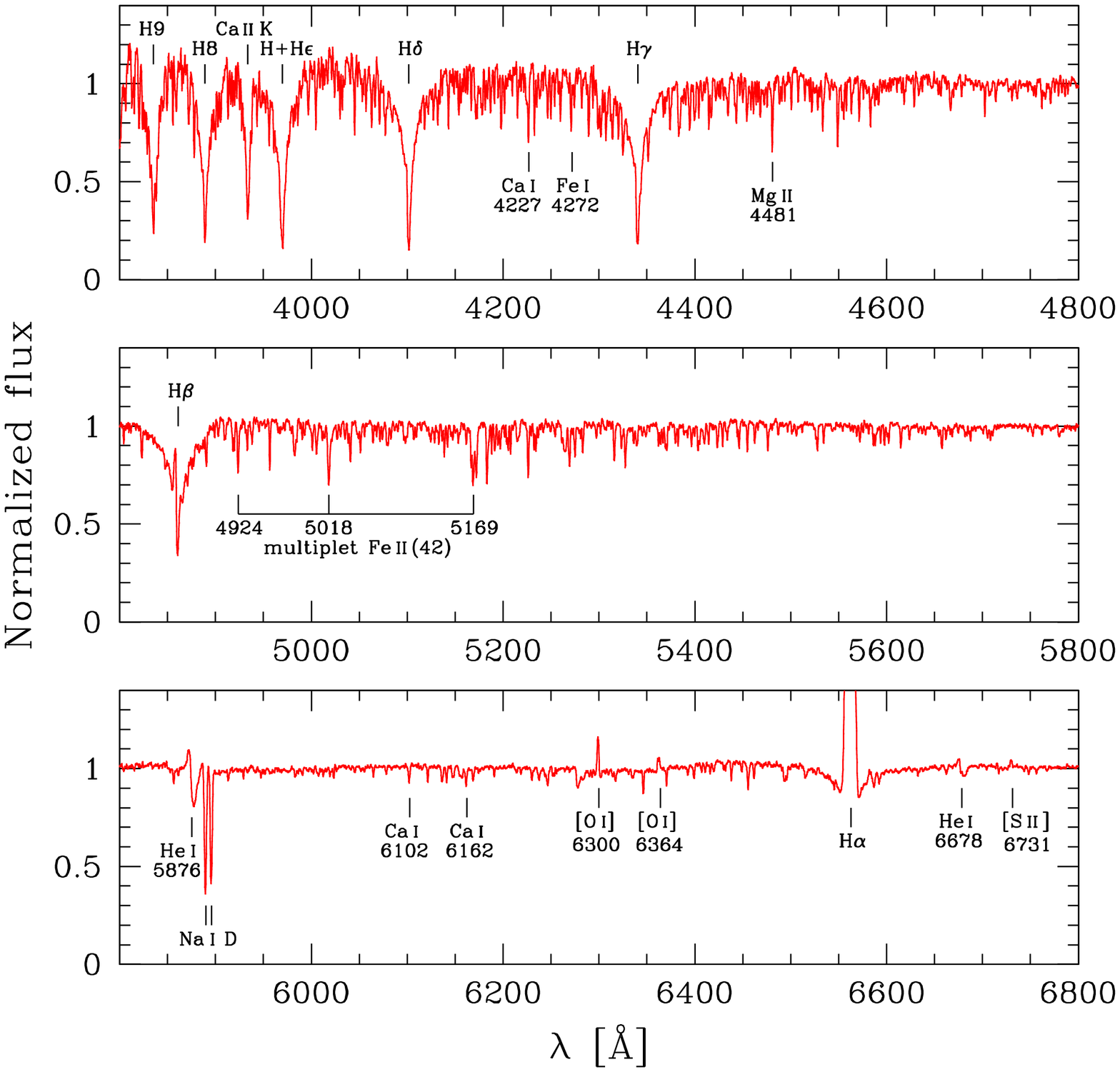}}
\FigCap{Normalized spectrum of V735 Sgr in the range 3800--6800 \AA.
Some lines observed in young stellar objects of spectral type A
are marked. Note the presence of [O {\scriptsize I}] and
[S {\scriptsize II}] lines in emission.}
\end{figure}

In Fig.~7, we show four lines with asymmetric profiles: H$\alpha$, H$\beta$,
He {\scriptsize I} 5876, and He {\scriptsize I} 6678. The H$\alpha$ profile
has two peaks with the secondary peak located blueward of the primary one
and reaching 87\% of its intensity. Such a two-component profile
stems from the presence of a circumstellar accretion disk observed at large
inclination angle (almost edge on). According to the work by Grinin and
Rostopchina (1996) the double-peaked H$\alpha$ line is characteristic
for active Herbig Ae/Be stars, that is intermediate-mass stars showing
rapid, large-amplitude, irregular brightness variations caused by dust clouds
orbiting the central star in the disk. Similar H$\alpha$ profile is observed
in known UX Ori-type variables such as HK Ori, VV Ser, WW Vul (Reipurth \etal 1996).
The absorption line of H$\beta$ in V735 Sgr has a blueshifted central emission.
The two lines of He {\scriptsize I} have the inverse P Cygni profile
indicating mass inflows onto the central star.

\begin{figure}[htb!]
\centerline{\includegraphics[angle=0,width=124mm]{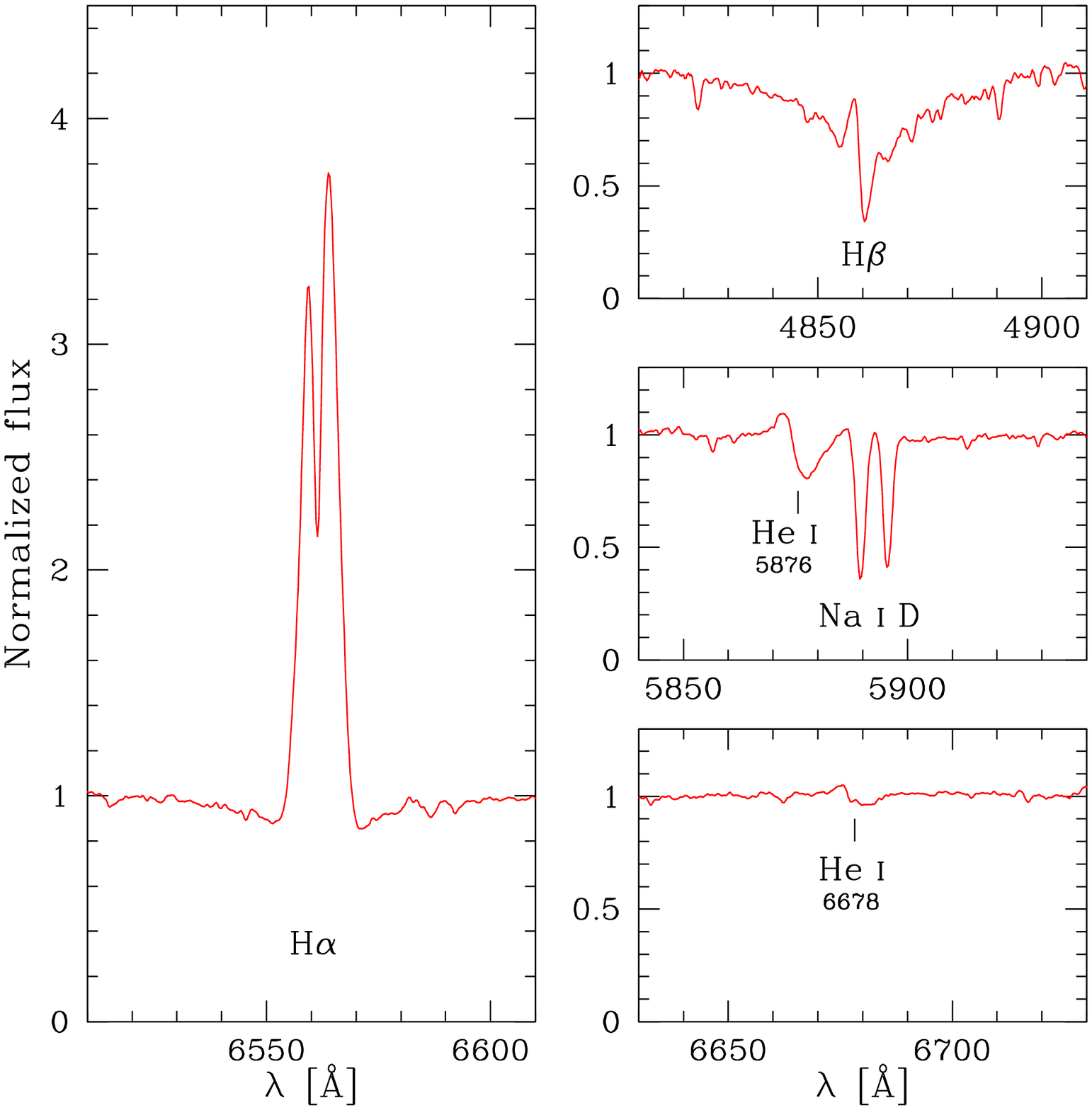}}
\FigCap{Profiles of four dramatically distorted lines, H$\alpha$, H$\beta$,
He {\scriptsize I} 5876, and He {\scriptsize I} 6678, due to disk
accretion in V735 Sgr. The wavelength and intensity are to scale in all
panels.}
\end{figure}

Determination of effective temperature in young stellar objects is complicated.
Application of the standard spectral classification from normal stars
has to be treated with caution (Gray and Corbally 2009).
The measured equivalent widths of $W$(Ca {\scriptsize II} K)$=5.2$\AA,
$W$(Ca {\scriptsize II} H + H$\epsilon$)$=12.5$\AA,
$W$(H$\delta$)$=13.7$\AA~and the presence of emission features indicate
type A3e for V735 Sgr. However, the real spectral type of the underlying
star may be slightly different, since the Ca {\scriptsize II} K line
is partially formed in the circumstellar disk and thus it is variable in time.
Helium lines and high order Balmer lines can also be affected by disk
absorption. Precise temperature determination requires a dedicated spectral
variability campaign that would help to identify constant photospheric
features (usually singly-ionized metals) and to compare them with
standards as proposed by Gray and Corbally (1998)
or Mora \etal (2001), for instance.


\Section{Conclusion}

Based on long-term OGLE photometry, brightness measurements in infrared
bands, and a moderate-resolution optical spectrum we showed that, so far
mysterious, rapidly varying star V735 Sgr is an active young stellar object
of Herbig Ae/Be type. This classification is confirmed by irregular
3-mag variations (of UX Ori type), the early spectral type of A3e and
a significant infrared excess, though the star does not seem to be associated
with a diffuse nebula and the absorption line Li {\scriptsize I} 6707
is absent in the spectrum. The OGLE light curves, flux-calibrated
and normalized spectra of V735 Sgr are available at
\begin{center}
{\it ftp://ftp.astrouw.edu.pl/ogle/ogle4/V735Sgr/}
\end{center}


\Acknow{
We thank OGLE observers for their contribution to the collection
of the photometric data over the years. The OGLE project
has received funding from the National Science Centre,
Poland (grant number MAESTRO 2014/14/A/ST9/00121 to A.U.).
R.A. acknowledges financial support from the DIDULS Regular PR17142
by Universidad de La Serena.}


\end{document}